\documentclass[superscriptaddress,aps,prl,twocolumn,showpacs,nofootinbib,longbibliography,notitlepage,floatfix]{revtex4-1}
\usepackage{etex}
\usepackage{amsmath,amssymb,amsthm}
\usepackage[colorlinks=true,citecolor=blue,urlcolor=blue]{hyperref}
\usepackage[pdftex]{graphicx}
\usepackage{times,txfonts}
\usepackage{braket}
\usepackage{color}
\usepackage{natbib}
\usepackage{amsmath,blkarray}
\usepackage{mathtools}
\usepackage{ulem}
\usepackage{latexsym}
\usepackage{tabularx, booktabs}
\usepackage{graphics,epstopdf}
\usepackage{graphicx}
\usepackage{float}
\usepackage{romannum}
\usepackage{amsfonts}
\usepackage{color,soul}

\newcommand{\be}{\begin{equation}}
\newcommand{\ee}{\end{equation}}
\newcommand{\ba}{\begin{eqnarray}}
\newcommand{\ea}{\end{eqnarray}}

\usepackage{multirow}
\usepackage{appendix}
\usepackage{url}

\begin{document}

\title{Digitized Adiabatic Quantum Factorization} 

\author{Narendra N. Hegade}
\email{narendrahegade5@gmail.com}
\affiliation{International Center of Quantum Artificial Intelligence for Science and Technology~(QuArtist) \\ and Physics Department, Shanghai University, 200444 Shanghai, China}

\author{Koushik Paul}
\email{koushikpal09@gmail.com}
\affiliation{International Center of Quantum Artificial Intelligence for Science and Technology~(QuArtist) \\ and Physics Department, Shanghai University, 200444 Shanghai, China}

\author{F. Albarr\'an-Arriagada}
\email{pancho.albarran@gmail.com}
\affiliation{International Center of Quantum Artificial Intelligence for Science and Technology~(QuArtist) \\ and Physics Department, Shanghai University, 200444 Shanghai, China}

\author{Xi Chen}
\email{xi.chen@ehu.eus}
\affiliation{Department of Physical Chemistry, University of the Basque Country UPV/EHU, Apartado 644, 48080 Bilbao, Spain}

\author{Enrique Solano}
\email{enr.solano@gmail.com}
\affiliation{International Center of Quantum Artificial Intelligence for Science and Technology~(QuArtist) \\ and Physics Department, Shanghai University, 200444 Shanghai, China}
\affiliation{Department of Physical Chemistry, University of the Basque Country UPV/EHU, Apartado 644, 48080 Bilbao, Spain}
\affiliation{IKERBASQUE, Basque Foundation for Science, Plaza Euskadi 5, 48009 Bilbao, Spain}
\affiliation{Kipu Quantum, Kurwenalstrasse 1, 80804 Munich, Germany}

\date{\today}

\begin{abstract}
Quantum integer factorization is a potential quantum computing solution that may revolutionize cryptography. Nevertheless, a scalable and efficient quantum algorithm for noisy intermediate-scale quantum computers looks far-fetched. We propose an alternative factorization method, within the digitized-adiabatic quantum computing paradigm, by digitizing an adiabatic quantum factorization algorithm enhanced by shortcuts to adiabaticity techniques. We find that this fast factorization algorithm is suitable for available gate-based quantum computers. We test our quantum algorithm in an IBM quantum computer with up to six qubits, surpassing the performance of the more commonly used factorization algorithms on the long way towards quantum advantage.
\end{abstract}

\maketitle

{\it Introduction.--} Quantum computers have the potential to solve certain computational problems significantly faster compared to classical computers. One remarkable example is the integer factorization problem, where no classical algorithms are known to have a polynomial-time solution. In 1994, P. Shor proposed a quantum algorithm to solve the integer factorization problem in polynomial time on a universal gate-based quantum computer  \cite{shor1994algorithms, shor1999polynomial}. However, we need a fault-tolerant quantum computer for reaching usefulness, which is far from the current noisy intermediate-scale quantum (NISQ) devices. Despite many demonstrations, the biggest number factored so far on an actual quantum computer using Shor's algorithm does not go beyond two digits \cite{lu2007demonstration, lanyon2007experimental, martin2012experimental, amico2019experimental}. An alternative paradigm to implement factorization algorithms is adiabatic quantum computation (AdQC), which is polynomially equivalent to the gate-based model \cite{AharonovSIAMRev2008}. In this paradigm, we map the factorization problem to an optimization one \cite{burges2002factoring}, which has been implemented in several architectures \cite{peng2008quantum, xu2012quantum, xu2017experimental,jiang2018quantum, dridi2017prime}. Nevertheless, due to low coherence time of quantum systems compared with the running time of algorithms in AdQC, its advantage is not clear~\cite{Gibney2017Nature}. In 2016, Barends \textit{et al.} \cite{barends2016digitized} used a digital quantum computer to implement a quantum adiabatic algorithm (QAA) by digitizing the adiabatic evolution. However, the large number of gates needed makes it still impractical for useful applications in NISQ devices.

In general, the efficiency of an algorithm is given by the run time to perform the computation. In the gate model, this run time depends on the number of quantum gates required by the algorithm. According to the adiabatic theorem, for the digital implementation of QAA, the number of gates depends on the minimum gap in the Hamiltonian spectrum. Therefore, the circuit depth increases rapidly when the minimum energy gap decrease for the digital version of QAA. In such scenario, shortcuts to adiabaticity (STA) techniques \cite{torrontegui2013shortcuts, del2013shortcuts, guery2019shortcuts, deffner2014classical} can help us to mimic fast adiabatic evolution by suppressing the non-adiabatic transitions \cite{demirplack1, demirplack2, berry, del2012assisted}. In this context, counterdiabatic (CD) driving is one of the most general techniques of STA. Despite its advantages, the exact CD driving is hard to calculate, and its nonlocal nature makes it cumbersome for applications. Instead, one can consider approximate CD driving protocols \cite{sels2017minimizing,claeys2019floquet}, which do not require the knowledge of the Hamiltonian spectrum for its calculation and can easily be implemented experimentally. It has been shown that, even with local CD terms, a drastic improvement in the fidelity can be obtained for many-body systems \cite{saberi2014adiabatic,campbell2015shortcut,hartmann2020many,prielinger2020diabatic,passarelli2020counterdiabatic}.   

Recently, Hegade \textit{et al.}~\cite{hegade2021shortcuts} showed the advantage of considering CD terms in the digital version of QAA, which drastically decreases the circuit depth for implementing quantum algorithms. This new paradigm called digitized-adiabatic quantum computing (DAdQC) opens the door to implement efficient algorithms for the NISQ era. In this work, we propose an integer factorization algorithm in the DAdQC paradigm, obtaining a reduction in the number of quantum gates and improved performance, making it suitable for NISQ devices. We use two different approaches for the factorization problem: the direct optimization method and the binary multiplication table method combined with classical preprocessing. Moreover, we test our algorithm in the IBM quantum computer with up to six qubits, obtaining better fidelities in all the cases under study compared to the most popular factorization algorithms. This work shows the potential of DAdQC to advance the field towards the goal of quantum advantage for practical applications with current technology.

{\it Approach 1: Direct optimization.---} Consider $N$ to be an integer number with $p$ and $q$ being its prime factors, so that $N-pq = 0$. Now, we can define the cost function as,
\begin{equation}
    f(x,y) = (N - xy)^2,
\end{equation}
with $x,y\in \mathbb{Z}^+$. As $f(x,y) \geq 0$, the minimum of the function ($f(x_s, y_s) = 0$) is reached only if $x_sy_s=N$, obtaining the solution of the factorization problem. Without losing generality, we will assume that $N$ is an odd integer. It follows that the factors $x$ and $y$ must also be odd numbers. To solve this problem in a quantum computer, we need to represent the factors $x$ and $y$ as a string of qubits. The exact length of the factors is previously unknown. However, the number of qubits sufficient to represent the prime factors in binary form requires $n_x = m\left(\lfloor\sqrt{N}\rfloor_{o}\right)-1$, $n_y=m\left(\left\lfloor\frac{N}{3}\right\rfloor\right)-1$ qubits \cite{peng2008quantum}. Here, $\lfloor a\rfloor$ ($\lfloor a\rfloor_{o}$) denotes the greatest (odd) integer less than or equal to $a$,  while $m(b)$ indicates the smallest number of bits required for representing $b$. 

We encode the solution of the factorization problem in the ground state of a Hamiltonian,
\begin{equation}
    H_f = \left[NI- \left(\sum_{l=1}^{n_x} 2^{l} \hat{x}_{l} + I\right)\left(\sum_{m=1}^{n_y} 2^{m} \hat{y}_{m} + I\right)\right]^2,
    \label{optHamilt}
\end{equation}
where $\hat{x}_l = \frac{I - \sigma^z_l}{2}$, and $\hat{y}_m = \frac{I - \sigma^z_m}{2}$. This Hamiltonian can be written in a general form as
\begin{equation}
\begin{split}
H_{f}=& \sum_{i} \tilde{h}_{i}^z \sigma_{i}^{z} + \sum_{i<j} \tilde{J}_{i j} \sigma_{i}^{z} \sigma_{j}^{z} + \sum_{i<j<k} \tilde{K}_{i j k} \sigma_{i}^{z} \sigma_{j}^{z} \sigma_{k}^{z} \\& + \sum_{i<j<k<l} \tilde{L}_{i j k l} \sigma_{i}^{z} \sigma_{j}^{z} \sigma_{k}^{z} \sigma_{l}^{z},
\end{split}
\label{direct_Hamiltonian}
\end{equation}
where $\tilde{J}_{i j}$, $\tilde{K}_{i j k}$ and $\tilde{L}_{i j k l}$ are the two, three, and four-body interaction terms, respectively. To find the ground state of this Hamiltonian, we initialize the system in the $\ket{+}^{\otimes n}$ state, which corresponds to the ground state of the initial Hamiltonian $H_i = \sum_i \tilde{h}_x \sigma^x_i$, and evolve the system adiabatically to the ground state of $H_f$. The total Hamiltonian thus for the adiabatic evolution can be expressed as,
\begin{equation}
    H_{ad}(t) = (1-\lambda(t)) H_i + \lambda(t) H_f,
    \label{eqn4}
\end{equation}
where $\lambda(t)=\sin ^{2}\left[\frac{\pi}{2} \sin ^{2}\left(\frac{\pi t}{2 T}\right)\right]$ is the scheduling function such that, $\dot{\lambda}(t)$ and $\ddot{\lambda}(t)$ vanishes at the beginning and end of the protocol. For finding the ground state of $H_f$, especially when the number of qubits required is large, the algorithm's run-time will be extensive. Subsequently, to perform the adiabatic evolution, the required computational cost is significantly higher which makes it rather unrealistic for implementation of factoring big numbers on a NISQ devices. These challenges can be overcome by using STA techniques by adding an auxiliary CD term $H_{CD} = \dot{\lambda}(t) A_{\lambda}$ to the Hamiltonian, which helps to achieve fast evolution by suppressing the non-adiabatic transitions. Here $A_{\lambda}$ is known as adiabatic gauge potential \cite{sels2017minimizing, hatomura2020controlling}, and its calculation for a many-body system is a demanding task. Here, we follow the method proposed by Sels \textit{et al.}~\cite{sels2017minimizing} to calculate the approximate gauge potential using variational approach, where a local CD term has been introduced to enhance the performance of quantum annealing.

\begin{figure}
    \centering
    \includegraphics[width=\linewidth]{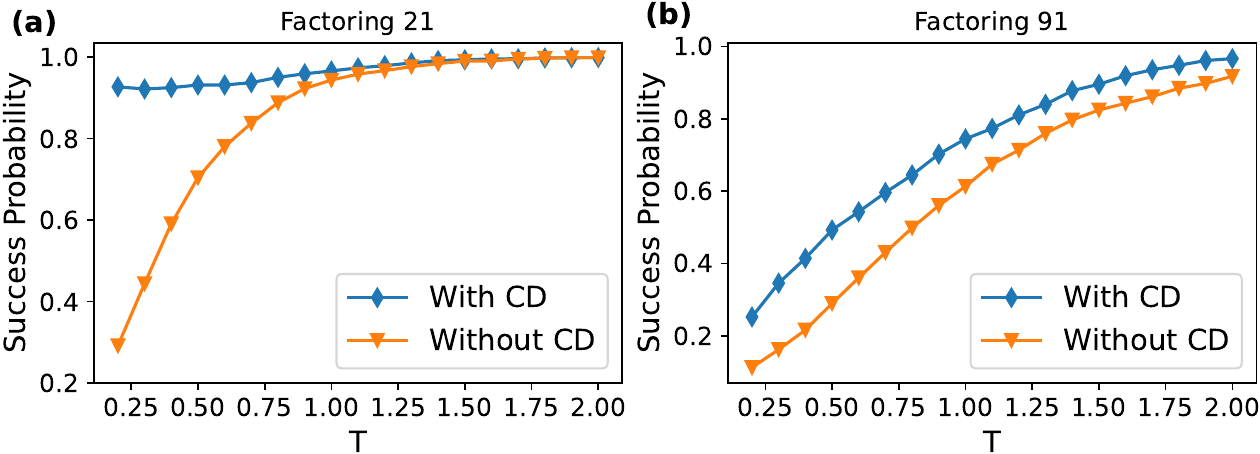}
    \caption{Ground state probability as a function of total evolution time for factorizing 21 and 91. The blue line corresponds to the evolution with local CD driving and the orange line is for without CD driving.}
    \label{fig1}
\end{figure}

For the Hamiltonian in Eq.~\eqref{eqn4}, the simplest form of the approximate CD term is an external magnetic field along the y-direction,
\begin{equation}
    \tilde{A}_{\lambda}=\sum_{j} \alpha_{j}(t) \sigma_{j}^{y}.
    \label{LCD}
\end{equation}
Here, the local CD coefficient $\alpha_{j}(t)$ is calculated as
\begin{equation}
    \alpha_j (t) = \frac{h_j^x(t) \dot{h}_j^z(t) - h_j^z(t) \dot{h}_j^x(t)}{R_j(t)},
    \label{lcd}
\end{equation}
where, $R_j(t) = 2 \dot{\lambda} ( {h_j^x}^2 + {h_j^z}^2 + 2 \sum_{k} J_{jk}^2 +3 \sum_{k<l} K_{jkl}^2 + 4 \sum_{k<l<m} $ $L_{jklm}^2 )$. Note that, in Eq.~\eqref{lcd}, the scheduling function has been incorporated in the new set of parameters, $h_j^z(t) = \lambda(t) \tilde{h}_j^z$, $h_j^x(t) = \lambda(t) \tilde{h}_j^x$ and so on (see supplementary material). The total Hamiltonian by including the CD term is $H(t)= H_{ad}(t) + \dot{\lambda}(t) \tilde{A}_{\lambda}$. This Hamiltonian can be written as sum of $K$ terms with at most 4-local interactions, i.e.,  $H(t)=\sum_{k=1}^{K} H_{k}(t)$. For the time evolution of the system, we discretize the continuous evolution $U(0, T) = \mathcal{T} \exp \{-i \int_{0}^{T} H(t) dt \}$ using the product formula \cite{poulin2011quantum},
\begin{equation}
    U(0, T) \approx \prod_{j=1}^{M} \prod_{k} \exp \left\{-i H_{k}(j \Delta t) \Delta t\right\},
    \label{evolution}
\end{equation}
where $M$ is the total number of discrete steps. Each term in the product can be efficiently implemented using a set of quantum gates.  In Fig.~\ref{fig1}, we compared the probability of obtaining the ground state measured in the computational basis for different evolution times $T$ with and without CD driving. 
As in Fig.~\ref{fig1} (a), the ground state corresponding to factorization of $21$ is $\ket{111}$, we can see that, even for very small total evolution time $T$, we can obtain $>90\%$ success probability by adding the CD term. In Fig.~\ref{fig1} (b), the ground state corresponding to factorizing the number $91$ is $\ket{11011}$, that represent the factors $q = (y_3 y_2 y_1 1) = (1101)=13$, and $p = (x_2x_11) = (111)=7$.

\begin{figure}
    \centering
    \includegraphics[width =0.9\linewidth]{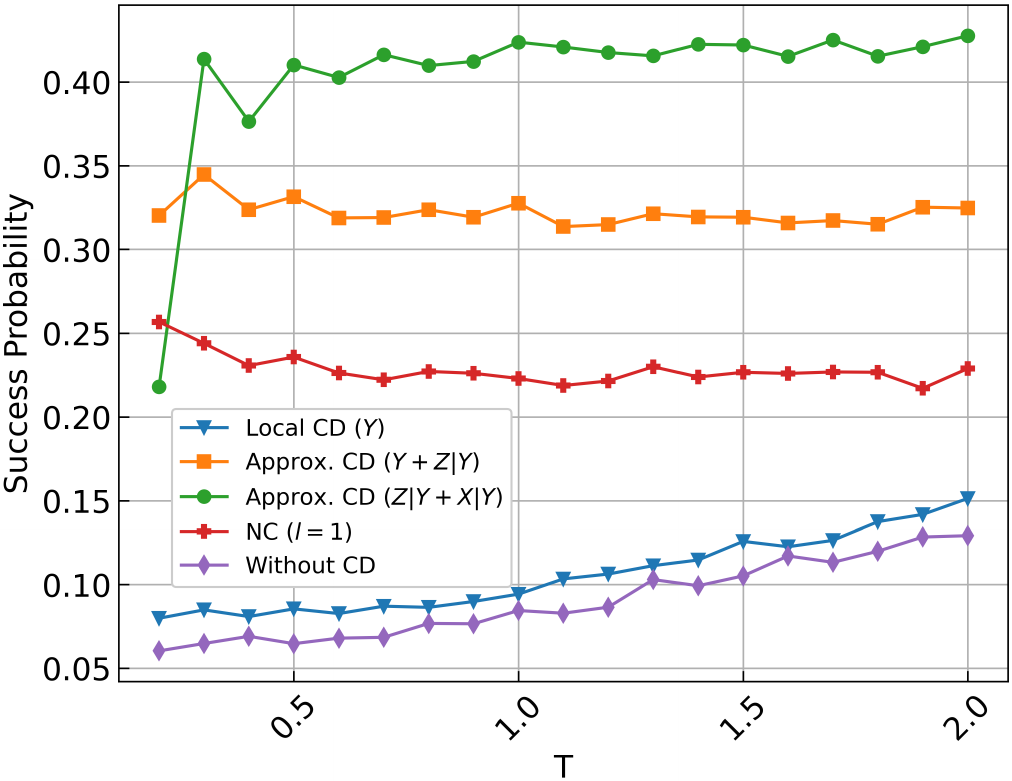}
    \caption{The success probability of obtaining the ground state $(\ket{0100})$ as a function of total evolution time for the Hamiltonian in Eq.~(\ref{eqn7}) corresponding to factoring the number $2479 = 37 \times 67$ using the CD term with multiple parameters from Eq.~\eqref{new_CD} (green and orange), the CD term obtained from NC ansatz in Eq.~\eqref{approx_cd} (red), the local CD in Eq.~\eqref{LCD} (blue) and the corresponding adiabatic case (purple) by simulating on \texttt{qasm} simulator.}
    \label{factor_2479}
\end{figure}

{\it Approach 2: Binary Multiplication Table.---}
For the Hamiltonian considered in Eq.~\eqref{direct_Hamiltonian}, the main disadvantage is that the spectral range (ratio of largest and smallest eigenvalue) increases exponentially with the system size, which makes it inefficient in terms of resources required for the computation since the gate complexity for implementing the Hamiltonian scales with the integrated norm $\int_{0}^{T}dt \|H_{ad}(t)\|$ \cite{berry2020time}. In order to reduce the complexity of the method, we consider the binary multiplication table to obtain the Hamiltonian using classical preprocessing \cite{burges2002factoring, schaller2007role, xu2012quantum, dattani2014quantum}. 
To illustrate this, we consider factoring the number 2479. The bit-lengths $n_x$ and $n_y$ for the factors $x$ and $y$ are chosen as 7 and 6, respectively. Since the prime numbers are odd, we set the first and last bit as 1. Table \Romannum{1}  (supplementary material) represents the multiplication table for factorizing 2479. Here, we introduced binary variables $c_{ij}$ as the carriers. By adding each column in the table, we obtain a set of simultaneous equations, called factorization equations. We simplify these equations further to reduce the total qubit requirement by applying classical preprocessing based on binary logical constraints. The time complexity of this simplification scales as $\mathcal{O}(n^3)$, i.e., polynomial in the bit length of the number being factored \cite{xu2012quantum}. The number of variables in the final equations decides the total qubit required for the computation. It has been observed that, after classical preprocessing, the number of qubits required for the factorization scales approximately as $\mathcal{O}(n)$ \cite{anschuetz2019variational, karamlou2020analyzing}, and without any simplification $\mathcal{O}(n \log n)$ qubits \cite{burges2002factoring}. Finally, by mapping the variables to the qubit operator, $q_i = (1-\sigma^z_i)/2$, the problem Hamiltonian can be constructed (see the supplementary for the detailed calculation). The Hamiltonian corresponding to factoring the number 2479 after classical preprocessing reads 
\begin{equation}
    \begin{split}
    H_f = & -2.5 \sigma_1^z - 1.5 \sigma_2^z + 0.75 \sigma_3^z -0.5 \sigma_4^z + 0.25 \sigma_1^z \sigma_2^z \\ & -1.5 \sigma_1^z \sigma_3^z - \sigma_1^z  \sigma_4^z + 0.5 \sigma_2^z  \sigma_3^z + 1.5 \sigma_2^z  \sigma_4^z\\ & + 0.5 \sigma_3^z \sigma_4^z+ 0.75 \sigma_1^z \sigma_2^z \sigma_3^z + 5.75 \: \mathbb{I} \, .
    \label{eqn7}
    \end{split}
\end{equation}
The local CD driving method, considered previously, generally gives better performance than the naive approach. However, the improvement is not always significant (see Fig.~\ref{fig1}b). The performance can be further improved by considering higher-order terms in the CD Hamiltonian. Recently, a systematic approach for constructing approximate CD driving was proposed by P. W. Claeys {\it et al.}~\cite{claeys2019floquet}, where the adiabatic gauge potential is chosen as
\begin{equation}
    A_{\lambda}^{(l)} = i \sum_{k = 1}^l \alpha_k(t) \underbrace{[H_{ad},[H_{ad},..., [H_{ad},}_{2k-1}\partial_{\lambda} H_{ad}]]] .
    \label{gauge}
\end{equation}
Here, $l$ corresponds to the expansion order, and when $l \to \infty$ we will get the exact gauge potential. By considering only the first-order expansion, we obtain the approximate CD term for the problem Hamiltonian in Eq.~\eqref{eqn7} as
\begin{equation}
    \begin{split}
    {A}_{\lambda}^{(1)}=& 2 \alpha_{1}(t) \tilde{h}_{x} \left[ \sum_{i} \tilde{h}_i^z \sigma^{y}_{i} + \sum_{i<j} \tilde{J}_{ij} \left(\sigma^{z}_{i} \sigma^{y}_{j}+\sigma^{y}_{i} \sigma^{z}_{j}\right) \right. \\& 
    \left. + \sum_{i<j<k} \tilde{K}_{ijk} \left(\sigma^{z}_{i} \sigma^{z}_{j}\sigma^{y}_{k} +\sigma^{z}_{i} \sigma^{y}_{j} \sigma^{z}_{k} +\sigma^{y}_{i} \sigma^{z}_{j} \sigma^{z}_{k}\right) \right]. 
    \end{split}
    \label{approx_cd}
\end{equation}
Using the variational method, we find the optimal CD coefficient $\alpha_{1}(t) =  0.0830/[h_x^2(1 - \lambda)^2 + 5.0112 \lambda^2]$. Note that Eq.~\eqref{approx_cd} represents a general form of the approximate CD term for an Ising spin chain that consists of a single CD coefficient. While this improves the success probability significantly, the performance can be enhanced further by introducing more coefficients that can be optimized using the variational method. In fact, we define an operator pool, restricted to only one and two spin terms that are obtained from the nested commutator (NC) expansion for $l=1$ and $l=2$ of Eq.~\eqref{gauge} , given by $\tilde{A}_{\lambda} = \{Y,\: Z|Y,\: X|Y \}$, where
\begin{align}
\begin{split}
Y = \sum_{i} \alpha_i(t) \sigma_{i}^{y}, \;\; Z|Y = \sum_{i<j} \beta_i(t) (\sigma_{i}^{z} \sigma_{j}^{y} + \sigma_{i}^{y} \sigma_{j}^{z})\\  X|Y =  \sum_{i<j} \gamma_i(t) (\sigma_{i}^{x} \sigma_{j}^{y} + \sigma_{i}^{y} \sigma_{j}^{x}).
\end{split}
\label{new_CD}
\end{align}
We assume that $\alpha_i(t) = \tilde{h}_i^z \alpha(t), \: \beta_i(t) = \tilde{J}_{ij}\beta(t), \: \gamma_i(t) = \tilde{J}_{ij}\gamma(t)$.  Fig.~\ref{factor_2479} shows the comparison of the success probability of obtaining the ground state using different CD terms, indicating that, introducing more parameters will increase the success probability. Also, we observe that with only three trotter steps, one can obtain the prime factors of 2479 with a high probability (see the probability distribution in the supplementary).

The cost of quantum adiabatic algorithm can be quantified by $C=T \max _{\lambda}\|H(\lambda(t))\|$ \cite{albash2018adiabatic}. In DAdQC, the total evolution time $T$ corresponds to the circuit depth, and the total cost corresponds to the total gate counts. Since the run time of the algorithm $T$ depends on the minimum energy gap $\Delta_{min}$, finding the time complexity of the adiabatic factorization algorithm with or without CD driving is a challenging task, and it is currently unknown. Present work verifies the applicability of the proposed method for factoring small numbers with few qubits. However, some shreds of evidence support that the inclusion of the CD term can be advantageous for factoring large numbers as well since the variational CD driving has been used to study a number of many-body systems, yielding significant improvement \cite{sels2017minimizing,hartmann2020many,prielinger2020diabatic,passarelli2020counterdiabatic, hegade2021shortcuts, hatomura2020controlling}. Moreover, the Hamiltonian corresponding to the factorization problem is stoquastic, and quantum Monte Carlo simulations can tackle such problems without facing any sign problem. It is believed that quantum adiabatic evolution or quantum annealing for stoquastic Hamiltonian might not give significant improvement over classical algorithms. However, the inclusion of the CD term makes the Hamiltonian non-stoquastic with imaginary entries at the off-diagonal terms, and no classical algorithms are known to implement such Hamiltonians efficiently. Also, there are several references showing that the inclusion of non-stoquastic drivers is advantageous \cite{vinci2017non, hormozi2017nonst, albash2019role}, and for some systems, it can give exponential speedup \cite{nishimori2017exponential}. Thus, the CD term can be considered as a type of non-stoquastic catalyst with the potential to lead to quantum speedups.
\begin{figure}
    \centering
    \includegraphics[width =\linewidth]{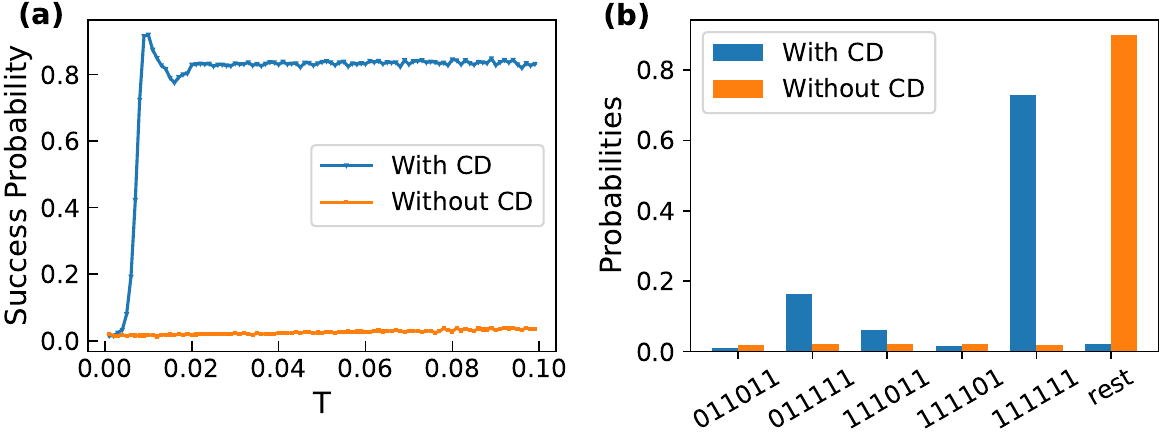}
    \caption{Factorization of $217 = 7 \times 31$: (a) The success probability as a function of total evolution time with and without CD driving using \texttt{qasm} simulator. (b) Probability distribution from 7-qubit quantum processor \texttt{ibmq\_casablanca}. Parameters chosen are: $T=0.01$, $\Delta t = 0.001$, $N_{shots} = 8192$.}
    \label{fig3}
\end{figure}

{\it Experimental analysis.---}For the experimental implementation of the algorithm, we consider IBM's cloud quantum computers \texttt{ibmq\_vigo} and \texttt{ibmq\_casablanca}, with 5-qubit and 7-qubit processors respectively. The first step in our simulation is to prepare the initial ground state $(\ket{0}+\ket{1})^{\otimes n}/\sqrt{2}$. By applying Hadamard gate on each qubit we prepare this initial state with a high fidelity (single qubit gate error $\sim10^{-4}$). For the time evolution of the system, we adopt the DAdQC paradigm by discretizing the total evolution time $T$ into a finite number of small steps of size $\Delta t = T/M$, where $M$ is the number of time steps. Using first-order Trotter-Suzuki formula, we approximate the evolution into a product of unitary operators corresponding to each time steps, that can be decomposed into a set of single qubit and two qubit gates (see supplementary material). Corresponding digitized time evolution operator is given in Eq.~\eqref{evolution}. For the first-order trotterization the error is $\mathcal{O}(\Delta t^2)$~\cite{suzuki1976generalized}. Also, since the Hamiltonian is time dependent, the step size $\Delta t$ must be smaller than the fluctuation time scale of the Hamiltonian, i.e., $\Delta t \ll\|\partial H / \partial t\|^{-1}$. An advantage of the gate model is that efficient circuit implementation for the digitized time evolution of a Hamiltonian with k-local interactions is well known \cite{raeisi2012quantum}. However, several sources of errors like gate error, readout error, and decoherence affect the outcome of the experiment. In order to obtain a better result, we consider readout error mitigation and circuit optimization techniques while implementing our algorithm on the actual hardware. In all our experiments, to obtain the probability distribution we consider the number of shots, $N_{shots} = 8192$.          

For the the direct optimization method, discussed earlier, we consider the factorization of the number $N=217 = 7 \times 31$, which requires at most $n=n_x + n_y -2 = 6$ ($n_x = 3$, $n_y= 5$) qubits to represent the unknown factors. The Hamiltonian that encodes the solution of the problem in its ground state is given in the supplementary material. In Fig.~\ref{fig3} (a), we compare the ground state success probability for the evolution with local CD driving and the naive approach. Fig.~\ref{fig3} (b), is the result obtained from the quantum processor \texttt{ibmq\_casablanca}. Here, we noticed that, although the original Hamiltonian contains several interaction terms, the local longitudinal fields are comparatively much stronger and dominate the evolution dynamics. Due to this, the local CD term can effectively suppress the transitions between the eigenstates. Subsequently, for the fast evolution, the CD terms become the dominant terms, and to avoid the device error, we neglected the interaction terms while implementing on the actual hardware. 

\begin{figure}
    \centering
    \includegraphics[width=0.9\linewidth]{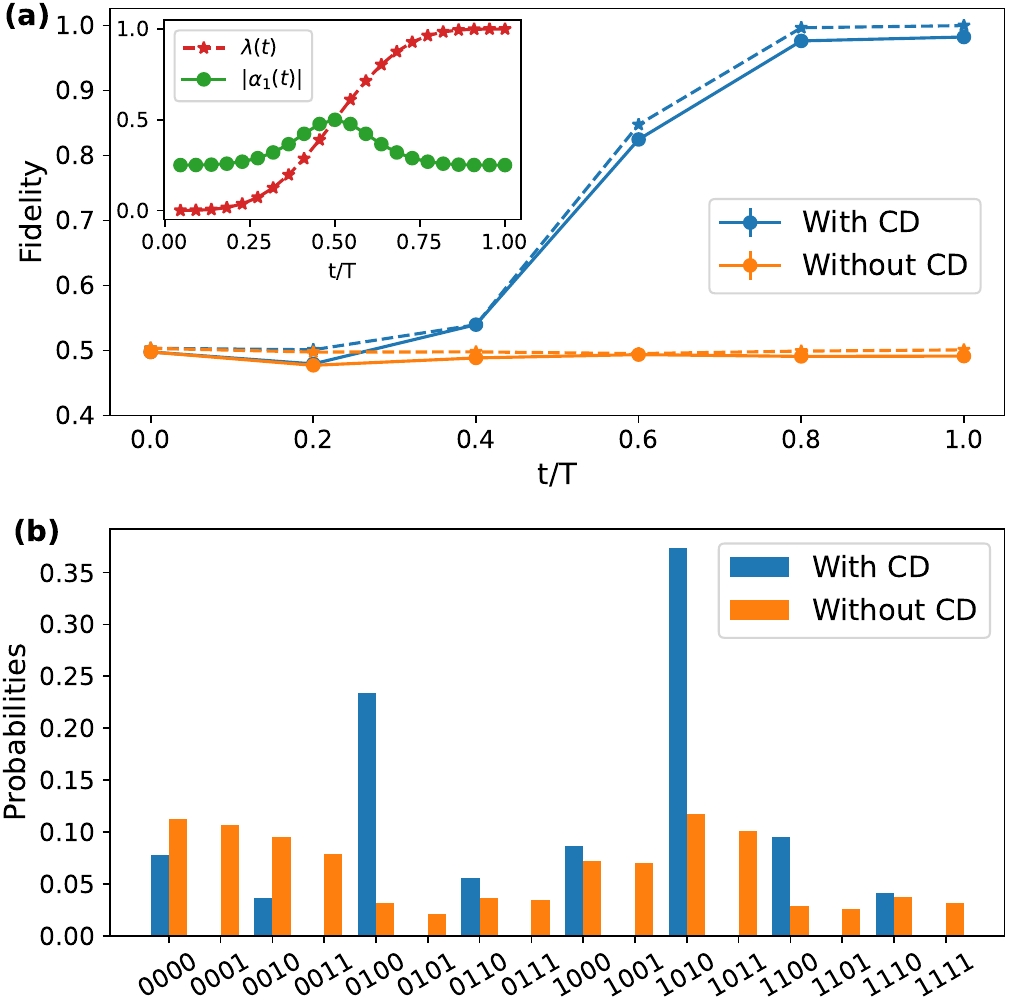}
    \caption{(a)The ground state fidelity as a function of evolution time for  both with and without CD. Inset: scheduling parameter $\lambda(t)$ and the variational CD coefficient $\alpha_1(t)$ as a function of time. Experimental parameters: $T=0.005$, $\Delta t=0.001$, $h_x=-2$. (b) Probability distribution for factoring 235 using 4-qubits on \texttt{ibmq\_casablanca} using only two trotter steps for $T=0.02$, $\Delta t=0.01$, $N_{shots} = 8192$.}
    \label{fig4}
\end{figure}

For the second approach, we consider factorization of 35 and 235 using binary multiplication table. After classical preprocessing, we obtain the problem Hamiltonian, that requires 2 qubits for factoring 35 and 4 qubits for factoring 235. We consider first order NC method for obtaining the approximate CD term which contains up to two spin interaction terms. The factors corresponding to 35 have same bit length, due to exchange symmetry the ground state is a two-fold degenerate state, i.e., $(\ket{01}+\ket{10})/\sqrt{2}$. Fig.~\ref{fig4} (a), depicts the fidelity versus evolution time for five trotter steps, where the blue line corresponds to the evolution with CD driving and orange line is for without CD driving. The solid line is for the experimentally obtained result from \texttt{ibmq\_vigo}, and the dotted line is the ideal simulation result. Even though five trotter steps are depicted here, one can obtain the final ground state with experimental fidelity 0.982 with only two trotter steps, with $T = 0.002$ and $dt = 0.001$. In Fig.~\ref{fig4} (b), the probability distribution obtained from \texttt{ibmq\_casablanca} for factoring the number 235 using two trotter steps are shown. Using CD  driving, the final ground state $\ket{0101}$ obtained has the highest success probability. In contrast, the adiabatic case requires hundreds of steps.

{\it Conclusions.---} We have proposed a method to speed up integer factorization using a digitized-adiabatic quantum algorithm, where the source adiabatic algorithm is enhanced via STA techniques with CD driving. We obtain a substantial improvement in the fidelities in a very short time, reducing the total gate required for the factorization problem compared to the conventional adiabatic evolution. We successfully factorized numbers bigger than those feasible using the Shor's algorithm with the same number of qubits, which makes it more suitable for NISQ devices~\cite{amico2019experimental, skosana2021demonstration}. Moreover, at variance with recent variational quantum factoring algorithm~\cite{anschuetz2019variational,karamlou2020analyzing}, our method does not require classical optimization and, therefore, does not face the problem of local minima or barren plateaus. Furthermore, the optimal local CD coefficient can be easily calculated, and the classical preprocessing used to reduce the number of qubits requires only polynomial time. 

We also show that by using more complex CD protocols, our results can be still improved. Therefore, a future direction is to go beyond current architectures with only nearest-neighbor interactions, in a co-design spirit, which can efficiently encode the problem Hamiltonian and the CD terms. In this context, recent works show an efficient alternative to implement k-local interactions  \cite{Ender2021arXiv, DriebShon2021arXiv, Fellner2021arXiv}, which could improve a variety of DAdQC protocols. Finally, this work shows how to build efficient quantum algorithms for NISQ devices, with the potential to approach a quantum advantage.

\begin{acknowledgments}
{\it Acknowledgments.} The authors are grateful to Adolfo del Campo for useful discussions. This work is supported by NSFC (12075145), STCSM (2019SHZDZX01-ZX04 and 20DZ2290900), SMAMR (2021-40), Program for Eastern Scholar, Basque Government IT986-16, Spanish Government PGC2018-095113-B-I00 (MCIU/AEI/FEDER, UE), projects QMiCS (820505) and OpenSuperQ (820363) of EU Flagship on Quantum Technologies, EU FET Open Grants Quromorphic (828826) and EPIQUS (899368). X. C. acknowledges the Ram\'on y Cajal program (RYC-2017-22482).
\end{acknowledgments}

\pagebreak
\onecolumngrid
\newpage

\begin{center}
\vspace{1cm}
\textbf{\large Supplementary material: Digitized-Adiabatic Quantum Factorization}
\end{center}

\setcounter{equation}{0}
\setcounter{figure}{0}
\setcounter{table}{0}
\setcounter{page}{1}
\makeatletter
\renewcommand{\theequation}{S\arabic{equation}}
\renewcommand{\thefigure}{S\arabic{figure}}
\renewcommand{\bibnumfmt}[1]{[S#1]}
\renewcommand{\citenumfont}[1]{S#1}

\section{Calculation of local CD term}

To factor a number $N$ in to its prime factors $x$ and $y$, we need $X=m\left(\lfloor\sqrt{N}\rfloor_{o}\right)-1$, $Y=m\left(\left\lfloor\frac{N}{3}\right\rfloor\right)-1$ qubits, where $\lfloor a\rfloor_{o}$ denotes the largest odd integer not larger than $a$,  while $m(b)$ denotes the smallest number of bits required for representing $b$. In binary notation $x=\left(x_{l_{1}-1} x_{l_{1}-2} \ldots x_{1} 1\right)$, $y=\left(y_{l_{1}+l_{2}-2} y_{l_{1}+l_{2}-3} \ldots y_{l_{1}} 1\right)$. So we can write

\begin{equation*}
    x=\sum_{i=1}^{l_{1}-1} 2^{i} x_{i}+1, \text { and } y=\sum_{j=l_{1}}^{l_{1}+l_{2}-2} 2^{j} y_{j}+1.    
\end{equation*}

Since both x and y are odd prime numbers, $x_0 = y_0 = 1$. This problem can be mapped to an optimization problem where the minimum of a function $f(x,y) = (N-x y)^2$ gives the solution of the factorization problem. It is possible to encode the solution of a minimization problem in the ground state of a Hamiltonian, 
\begin{equation}
    H_f = \left[NI- \left(\sum_{l=1}^{n_x} 2^{l} \hat{x}_{l} + I\right)\left(\sum_{m=1}^{n_y} 2^{m} \hat{y}_{m} + I\right)\right]^2,
\end{equation}
where $\hat{x}_l = \frac{I - \sigma^z_l}{2}$ and $\hat{y}_m = \frac{I - \sigma^z_m}{2}$. This Hamiltonian can be written in a general form as

\begin{equation}
H_{f}=\sum_{i} \tilde{h}_i \sigma_{i}^{z} + \sum_{i<j} \tilde{J}_{i j} \sigma_{i}^{z} \sigma_{j}^{z}+\sum_{i<j<k} \tilde{K}_{i j k} \sigma_{i}^{z} \sigma_{j}^{z} \sigma_{k}^{z}+\sum_{i<j<k<l} \tilde{L}_{i j k l} \sigma_{i}^{z} \sigma_{j}^{z} \sigma_{k}^{z} \sigma_{l}^{z},
\end{equation}
where $\tilde{J}_{i j}$, $\tilde{K}_{i j k}$ and $\tilde{L}_{i j k l}$ are the two, three, and four-body interaction terms, respectively. In order to find the ground state of this Hamiltonian, we follow the adiabatic theorem by preparing the ground state of an initial Hamiltonian $H_i = \sum_i \tilde{h}_i\sigma^x_i$ and evolve the system adiabatically to reach the final ground state of $H_f$. The total Hamiltonian for the adiabatic evolution is given by
\begin{equation}
    H = (1-\lambda) H_i + \lambda H_f \, .
\end{equation}

In order to speedup the slow adiabatic evolution and suppress the unwanted transitions, we add an extra term to the Hamiltonian called counter-diabatic (CD) driving term. The calculation and implementation of the exact CD term is not useful for the practical purpose, so we consider an approximate local CD driving proposed by Sels {\it et al.} (see Ref. [23] in the main text),
\begin{equation}
    \tilde{A}_{\lambda}=\sum_{j} \alpha_{j}(t) \sigma_{j}^{y}.
\end{equation}

For a specified control parameter $\lambda$, the CD term is given by $H_{CD} = \dot{\lambda}\tilde{A}_{\lambda}$, where $\tilde{A}_{\lambda}$ is the approximate gauge potential responsible for the non-adiabatic transitions and $\alpha_{j}(t)$ is the corresponding CD coefficient. For the optimal solution, we have to minimize the operator distance between the exact gauge potential and the approximate gauge potential, which is equivalent to minimizing the action,
\begin{equation}
{S_{\lambda}}\left(\tilde{A}_{\lambda}\right)=\operatorname{Tr}\left[G_{\lambda}^{2}\left(\tilde{A}_{\lambda}\right)\right],
\end{equation}
where the Hilbert-Schmidt norm $G_{\lambda}$ is given by
\begin{equation}
G_{\lambda}\left(\tilde{A}_{\lambda}\right)=\partial_{\lambda} H+i\left[\tilde{A}_{\lambda}, H\right].
\end{equation}

Moreover,
\begin{equation*}
    \partial_{\lambda} H = \frac{1}{\dot{\lambda}} \bigg(- \sum_i \dot{h}_i^x \sigma_i^x + \sum_i \dot{h}_i^z \sigma_i^z + \sum_{i <j} \dot{J}_{ij} \sigma_i^z \sigma_j^z + \sum_{i <j <k} \dot{K}_{ijk} \sigma_i^z \sigma_j^z \sigma_k^z + \sum_{i <j <k<l} \dot{L}_{ijkl} \sigma_i^z \sigma_j^z \sigma_k^z \sigma_l^z \bigg) \, ,
\end{equation*}
where, the scheduling function $\lambda(t)$ has been incorporated in the new set of parameters, $h_j^z(t) = \lambda(t) \tilde{h}_j^z$, $h_j^x(t) = \lambda(t) \tilde{h}_j^x$, $J_{ij}^z(t) = \lambda(t) \tilde{J}_{ij}^z$, and so on. Furthermore,
\begin{equation}
    \begin{split}
    G_\lambda =& \sum_i \left (- \frac{\dot{h}_i^x}{\dot{\lambda}} - 2 \alpha_i h_i^z \right) \sigma_i^x - 2 \sum_{i<j} \alpha_i J_{ij} \left(\sigma_i^x \sigma_j^z + \sigma_i^z \sigma_j^x \right) - 2 \sum_{i<j<k} \alpha_i K_{ijk} \left ( \sigma_i^x \sigma_j^z \sigma_k^z + \sigma_i^z \sigma_j^x \sigma_k^z + \sigma_i^z \sigma_j^z \sigma_k^x \right ) \\&
    -2 \sum_{i<j<k<l} \alpha_i L_{ijkl} \left( \sigma_i^x \sigma_j^z \sigma_k^z \sigma_l^z + \sigma_i^z \sigma_j^x \sigma_k^z \sigma_l^z + \sigma_i^z \sigma_j^z \sigma_k^x \sigma_l^z \right) + \sum_i  \left(\frac{\dot{h}_{i}^z}{\dot{\lambda}} - 2 \alpha_i h_i^x \right) \sigma_i^z + \sum_{i<j} \frac{\dot{J}_{ij}}{\dot{\lambda}} \sigma_i^z \sigma_j^z \\&
    + \sum_{i <j <k} \frac{\dot{K}_{ijk}}{\dot{\lambda}} \sigma_i^z \sigma_j^z \sigma_k^z + \sum_{i <j <k<l} \frac{\dot{L}_{ijkl}}{\dot{\lambda}}\sigma_i^z \sigma_j^z \sigma_k^z \sigma_l^z \, .
    \end{split}
\end{equation}
The action is calculated as
\begin{eqnarray}
  S &= & Tr(G_\lambda^2) \nonumber \\ & = & \sum_i (- \frac{\dot{h}_i^x}{\dot{\lambda}} - 2 \alpha_i h_i^z )^2 + 8 \sum_{i<j}  \alpha_i^2 J_{ij}^2 + 12 \sum_{i<j<k} \alpha_i^2 K_{ijk}^2 + 16 \sum_{i<j<k<l} \alpha_i^2 L_{ijkl}^2 + \sum_i (\frac{\dot{h}_i^z}{\dot{\lambda}} - 2 \alpha_i h_i^x )^2 + \sum_{i<j} \frac{ \dot{J}_{ij}^2}{\dot{\lambda}^2} + \sum_{i <j <k} \frac{\dot{K}_{ijk}^2}{\dot{\lambda}^2} + \sum_{i <j <k<l} \frac{\dot{L}_{ijkl}^2}{\dot{\lambda}^2} \, . \nonumber \\
\end{eqnarray}
By minimizing the action $\frac{\partial S}{\partial \alpha_i} = 0$, we will get the CD coefficient
\begin{equation}
    \alpha_i = \frac{h_i^x \dot{h}_i^z - h_i^z \dot{h}_i^x}{2 \dot{\lambda} \left( {h_i^x}^2 + {h_i^z}^2 + 2 \sum_{j} J_{ij}^2 +3 \sum_{j<k} K_{ijk}^2 + 4 \sum_{j<k<l} L_{ijkl}^2  \right)} \, .
\end{equation}
Therefore the local CD driving can be calculated as,
\begin{equation}
H_{CD}(t) = \dot{\lambda} \tilde{A}_{\lambda} = \sum_i \frac{h_i^x \dot{h}_i^z - h_i^z \dot{h}_i^x}{2 \left( {h_i^x}^2 + {h_i^z}^2 + 2 \sum_{j} J_{ij}^2 +3 \sum_{j<k} K_{ijk}^2 + 4 \sum_{j<k<l} L_{ijkl}^2  \right)} \, \sigma_i^y .
\end{equation}

\subsection{Example-1: quantum factorization of 217 using the direct optimization method}

For the adiabatic quantum factorization using direct optimization method, we consider the example of factoring the number $N=217 = 7 \times 31$, which requires $n_x = 3$, and $n_y= 5$ qubits to represent the factors. Hence, for the simulation we need total $(n_x-1) + (n_y-1) = 6$ qubits. The problem Hamiltonian can be calculated using Eq.~(2) from the main paper,
\begin{equation*}
    \begin{split} 
    H_f =& 16 \sigma^z_{1} \sigma^z_{2} \sigma^z_{3} \sigma^z_{4}+32 \sigma^z_{1} \sigma^z_{2} \sigma^z_{3} \sigma^z_{5}+64 \sigma^z_{1} \sigma^z_{2} \sigma^z_{3} \sigma^z_{6}-128 \sigma^z_{1} \sigma^z_{2} \sigma^z_{3}+64 \sigma^z_{1} \sigma^z_{2} \sigma^z_{4} \sigma^z_{5}+128 \sigma^z_{1} \sigma^z_{2} \sigma^z_{4} \sigma^z_{6}-256 \sigma^z_{1} \sigma^z_{2} \sigma^z_{4} \\& +256 \sigma^z_{1} \sigma^z_{2} \sigma^z_{5} \sigma^z_{6}-512 \sigma^z_{1} \sigma^z_{2} \sigma^z_{5}-1024 \sigma^z_{1} \sigma^z_{2} \sigma^z_{6} +1364 \sigma^z_{1} \sigma^z_{2}-32 \sigma^z_{1} \sigma^z_{3} \sigma^z_{4}-64 \sigma^z_{1} \sigma^z_{3} \sigma^z_{5}-128 \sigma^z_{1} \sigma^z_{3} \sigma^z_{6}-178 \sigma^z_{1} \sigma^z_{3} \\& -128 \sigma^z_{1} \sigma^z_{4} \sigma^z_{5} -256 \sigma^z_{1} \sigma^z_{4} \sigma^z_{6}-356 \sigma^z_{1} \sigma^z_{4}-512 \sigma^z_{1} \sigma^z_{5} \sigma^z_{6}-712 \sigma^z_{1} \sigma^z_{5}-1424 \sigma^z_{1} \sigma^z_{6} +4216 \sigma^z_{1}-64 \sigma^z_{2} \sigma^z_{3} \sigma^z_{4}-128 \sigma^z_{2} \sigma^z_{3} \sigma^z_{5}\\& -256 \sigma^z_{2} \sigma^z_{3} \sigma^z_{6}-356 \sigma^z_{2} \sigma^z_{3}-256 \sigma^z_{2} \sigma^z_{4} \sigma^z_{5}-512 \sigma^z_{2} \sigma^z_{4} \sigma^z_{6}-712 \sigma^z_{2} \sigma^z_{4}-1024 \sigma^z_{2} \sigma^z_{5} \sigma^z_{6}-1424 \sigma^z_{2} \sigma^z_{5}-2848 \sigma^z_{2} \sigma^z_{6} +8432 \sigma^z_{2} \\& +84 \sigma^z_{3} \sigma^z_{4}+168 \sigma^z_{3} \sigma^z_{5}+336 \sigma^z_{3} \sigma^z_{6}+1064 \sigma^z_{3}+336 \sigma^z_{4} \sigma^z_{5}+672 \sigma^z_{4} \sigma^z_{6}+2128 \sigma^z_{4}+1344 \sigma^z_{5} \sigma^z_{6}+4256 \sigma^z_{5}+8512 \sigma^z_{6}+26474 I
    \end{split}
\end{equation*}

The ground state of this Hamiltonian is $\ket{111111}$, which encodes the solution of the factorization problem $217=7 \times 31 = \mathbf{\underline{11}}1 \times \mathbf{\underline{1111}}1$. To find the ground state, we use the 7-qubit quantum processor \texttt{ibmq\_casablanca}, and the experimentally obtained result is shown in Fig. 3 (b), in the main text.       

\begin{figure}
    \centering
    \includegraphics[width=0.9\linewidth]{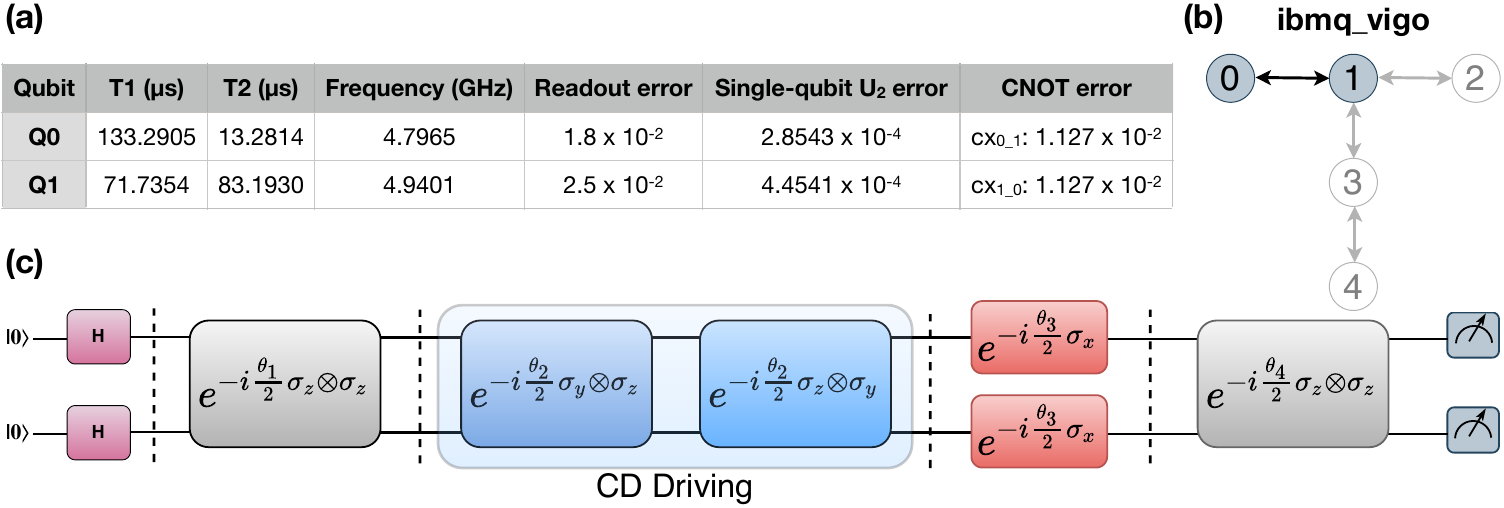}
    \caption{(a) Experimental parameters of the quantum processor \texttt{ibmq\_vigo}, (b) device layout, (c) circuit implementation for the time evolution of the Hamiltonian to factorize $35=7\times5$ using CD driving (two trotter steps).}
    \label{fig_device}
\end{figure}

\begin{figure}
    \centering
    \includegraphics[width =1 \linewidth]{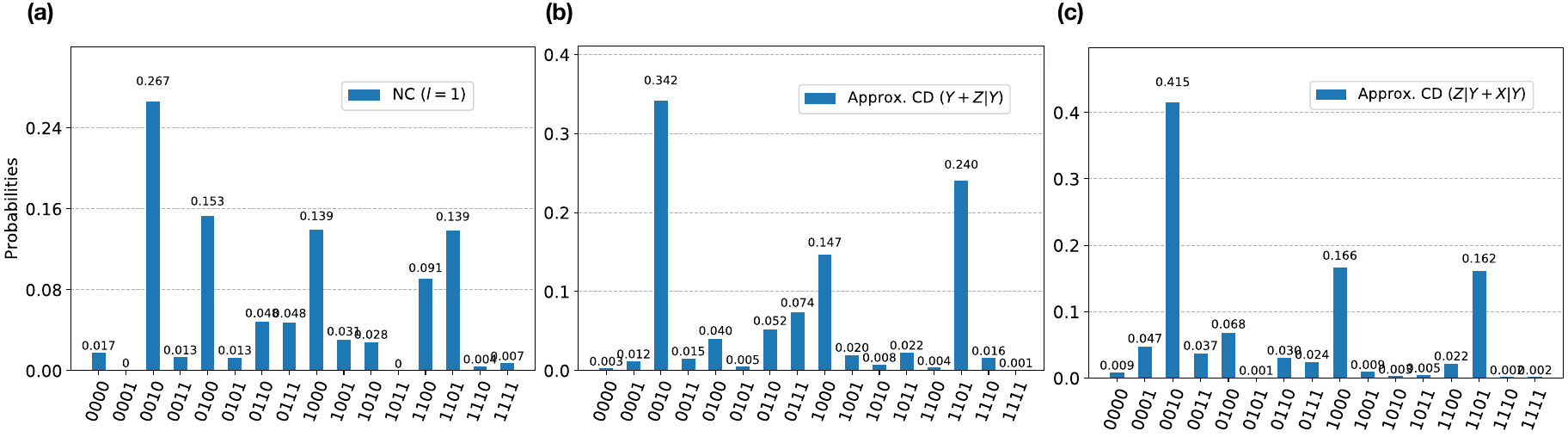}
    \caption{Probability distribution corresponding to the factorization of $2479 = 67 \times 37$ using three trotter steps obtained by considering three different CD terms. In all the cases the ground state $\ket{0100}$ (read from right to left) has the highest probability. The simulation parameters are: $\Delta t = 0.1$, $T=0.3$, and $N_{shots} = 10000$. }
    \label{probdist_2479}
\end{figure}

\subsection{Example-2: quantum factorization of $35=7\times5$}
To factorize the number 35 on a quantum processor, we consider the binary multiplication table method. After classical preprocessing, the number of qubits required for the factorization reduces to 2. Since the bit length of the factors 7 and 5 are same, the ground state is two fold degenerate due to exchange symmetry. The device layout, parameters and the quantum circuit for the digitized-adiabatic evolution using CD driving is shown in Fig.~\ref{fig_device}.      

\subsection{Example-3: quantum factorization of 2479 using the binary multiplication table} The binary multiplication table for factoring 2479 is given in Table~\ref{tab1}. For simplicity, we assumed that the bit length of the factors are known, and we consider $n_x = 7$, and $n_y = 6$. We set the first and last bit of the factors to be 1. Adding each column leads to a set of factorization equations. To reduce the number of qubits we apply the classical preprocessing scheme based on a binary logical constraints. we get the final set of equations as 
\begin{equation*}
    \begin{split}
        x_3 y_1 - y_1 =& \;0 \\  
        x_3 y_2 - y_1 =& \;0  \\
        x_3 + y_2 + c_{7,8} - 1 =& \;0 \\
        y_1 - y_2 - 2 c_{7,8} + 1 =& \;0 \\
        x_3 - 2y_1y_2 - y_1 + y_2 - 1 =& \;0. 
    \end{split}
\end{equation*}

By squaring and summing all the equations, we get the cost function as

\begin{equation}
    f(x,y,c) = (x_3 y_1 - y_1)^2 + (x_3 y_2 - y_1)^2 + (x_3 + y_2 + c_{7,8} - 1)^2 + (y_1 - y_2 - 2 c_{7,8} + 1)^2  + (x_3 - 2y_1y_2 - y_1 + y_2 - 1)^2.
\end{equation}

The minimum of this cost function $f_{min}(x,y,c) = 0$. By mapping the binary variables to the qubit operator, we obtained the final Hamiltonian given in Eq. 8 in the main manuscript. The probability distribution obtained at the end of the evolution by considering different CD terms is shown in Fig. \ref{probdist_2479}. In all the cases the ground state $\ket{0100}$ is obtained with highest success probability with only three trotter steps. 

\begin{table}
    \caption{ Multiplication table for $67 \times 37 = 2479$ in binary.}
    \resizebox{0.5\linewidth}{!}{
    \begin{tabular}{lllllllllllllllll}
        \hline
        \hline
         &&&&$2^{11}$&$2^{10}$&$2^{9}$&$2^{8}$&  $2^{7}$ & $2^{6}$ & $2^{5}$ & $2^{4}$ & $2^{3}$ & $2^{2}$ & $2^{1}$ & $2^{0}$ \\
         x&&&&&&&&&1& $x_5$&$x_4$& $x_3$ & $x_2$ & $x_1$ &1\\
         y&&&&&&&&&&1&$y_4$&$y_3$& $y_2$ & $y_1$ &1\\
         \hline
         &&&&&&&&&1&$x_5$& $x_4$ & $x_3$ & $x_2$ & $x_1$ & $1$ \\
         &&&&&&&&$y_1$&$y_1x_5$& $y_1x_4$ & $y_1x_3$ & $y_1x_2$ & $y_1x_1$ & $y_1$ &\\
         &&&&&&&$y_2$&$y_2x_5$& $y_2x_4$ & $y_2x_3$ & $y_2x_2$ & $y_2x_1$ & $y_2$ &&\\
         &&&&&&$y_3$&$y_3x_5$& $y_3x_4$ & $y_3x_3$ & $y_3x_2$ & $y_3x_1$ & $y_3$ &&&\\
         &&&&&$y_4$&$y_4x_5$& $y_4x_4$ & $y_4x_3$ & $y_4x_2$ & $y_4x_1$ & $y_4$ &&&&\\
         &&&&1& $x_5$&$x_4$& $x_3$ & $x_2$ & $x_1$ &1&&&&&\\
         carries &&&&$c_{10,11}$&$c_{9,10}$&$c_{8,9}$&$c_{7,8}$& $c_{6,7}$&$c_{5,6}$&$c_{4,5}$ & $c_{3,4}$ & $c_{2,3}$ & $c_{1,2}$ &&\\
         &&&&$c_{9,11}$&$c_{8,10}$&$c_{7,9}$& $c_{6,8}$&$c_{5,7}$&$c_{4,6}$ & $c_{3,5}$ & $c_{2,4}$&&&&\\
         \hline
         $x \times y = 2479$&&&&1&0&0&1&1&0&1&0&1&1&1&1 \\
         \hline
    \end{tabular}
    }
    \label{tab1}
\end{table}

\end{document}